\documentclass{aa}  
\usepackage{graphicx}
\usepackage{natbib}
\usepackage{hyperref}
\usepackage[varg]{txfonts}
\usepackage{xcolor}
\usepackage{float}
\usepackage{placeins}
\usepackage{stfloats}

\begin{document} 

  \title{Connecting integrated RGB mass loss from asteroseismology and globular clusters.}
\titlerunning{Connecting RGB mass loss from asteroseismology and globular clusters}

  \author{K. Brogaard
          \inst{\ref{difa}, \ref{aarhus}}
        A. Miglio \inst{\ref{difa},\ref{oas}} \and
        W. E. van Rossem\inst{\ref{difa}} \and
        E. Willett\inst{\ref{bhm}} \and 
        J. S. Thomsen\inst{\ref{difa}, \ref{aarhus}, \ref{oas}}             
 }

  \institute{
  {Department of Physics \& Astronomy, University of Bologna, Via Gobetti 93/2, 40129 Bologna, Italy}\label{difa}
  \and
  {Stellar Astrophysics Centre, Department of Physics \& Astronomy, Aarhus University, Ny Munkegade 120, 8000 Aarhus C, Denmark}\label{aarhus}
  \and
{INAF – Osservatorio di Astrofisica e Scienza dello Spazio, Via P. Gobetti 93/3, 40129 Bologna, Italy} \label{oas}
\and
{School of Physics and Astronomy, University of Birmingham, Edgbaston, Birmingham, B15 2TT, UK}
\label{bhm}
    }

  \date{Received XXX / Accepted XXX}

  \abstract{
    Asteroseismic investigations of solar-like oscillations in giant stars enable the derivation of their masses and radii. For mono-age mono-metallicity populations of stars this allows the integrated red giant branch (RGB) mass loss to be estimated by comparing the median mass of the low-luminosity RGB stars to that of the helium-core-burning stars (HeCB). 
}{
    We aim to exploit quasi mono-age mono-metallicity populations of field stars in the $\alpha$-rich sequence of the Milky Way (MW) to derive the integrated mass loss and its dependence on metallicity.
    By comparing to metal-rich globular clusters (GCs), we wish to determine whether the RGB mass loss differs in the two environments. 
}{
    Catalogues of asteroseismic parameters based on time-series photometry from the \textit{Kepler} and K2 missions cross-matched to spectroscopic information from APOGEE-DR17, photometry from 2MASS, parallaxes from Gaia DR3 and reddening maps are utilised.
    The RGB mass loss is determined by comparing mass distributions of RGB and HeCB stars in three metallicity bins. For two GCs, the mass loss is derived from colour-magnitude diagrams.
}{
    Integrated RGB mass loss is found to increase with decreasing metallicity and/or mass in the [Fe/H] range from $-0.9$ to $+0.0$. At [Fe/H]=$-0.50$ the RGB mass loss of MW $\alpha$-rich field stars is compatible with that in GCs of the same metallicity.  
}{   
    We provide novel empirical determinations of the integrated mass loss connecting field stars and GC stars at comparable metallicities. These show that mass loss cannot be accurately described by a Reimers mass-loss law with a single value of $\eta$. This should encourage further theoretical developments aimed at gaining a deeper understanding of the processes involved in mass loss.
    }

\maketitle

\section{Introduction}
Mass loss occurring along the red giant branch (RGB) is a poorly understood process that limits our ability to infer accurate ages of stars in subsequent phases, especially in the red-clump (RC, e.g., see \citealt{Casagrande2016, Anders2017}), to predict the dynamical evolution of planetary systems (e.g., see \citealt{Schroeder2008}), to determine the physical parameters shaping the horizontal branch in globular clusters (GCs, e.g., see \citealt{DAntona2002}), and to understand the formation channels of sdB stars, which affect the UV excess in old stellar systems (e.g., see \citealt{Han2002}).

In stellar models RGB mass loss is typically described by simple prescriptions \citep[e.g][]{Reimers1975b,Reimers1975a,SC2005}, with limited understanding of the underlying physical mechanism. 
The \textit{Kepler} \citep{Borucki2010} and K2 \citep{Howell2014} missions allowed asteroseismology of solar-like oscillations for giant stars in a few open and GCs. For these clusters, the average integrated RGB mass loss could therefore be measured by comparing the average mass of stars in the helium-core-burning (HeCB) phase (also known as the RC or red horizontal-branch (RHB) depending on the mass and metallicity) to the mass of the stars on the low-luminosity RGB (i.e. with $L$ < $L$(HeCB)) \citep{Miglio2012,Stello2016,Handberg2017,Tailo2022, Howell2022}. 
The numbers derived suggest a lower mass-loss efficiency in the open clusters compared to the GCs. 
\citet{Miglio2021} used the same method while exploiting the narrow mass distribution of \textit{Kepler} $\alpha$-rich field stars to derive an average integrated RGB mass loss in agreement with the open clusters.

In this letter, we determine average integrated mass loss for larger sub-samples of $\alpha$-rich field giants across different metallicity bins. These stars exhibit narrow mass distributions, allowing us to approximate the mono-age, mono-metallicity populations typical of clusters, while covering a broader range of metallicities and older ages. In the overlapping metallicity range, we also derive integrated mass loss from colour-magnitude diagrams (CMDs) of metal-rich GCs and show that the results are consistent with those from the field stars.

\section{Datasets}
\label{sec:observations}

We use the catalogues by Willett et al. (in prep.) based on asteroseismic measurements from \textit{Kepler} and K2 data combined with parallaxes from Gaia DR3 \citep{GaiaDR3-2022}, $K_s$ photometry from 2MASS \citep{Skrutskie2006}, and spectroscopic data from APOGEE DR17 \citep{APOGEEDR17-2022}.
We supplemented these catalogues with asteroseismic measurements by \citet{Elsworth2020} and Gaia DR3 zeropoint corrections from \citet{Lindegren2021} and \citet{Khan2023}. 

For the GCs, we used Hubble Space Telescope (HST) photometry from the HST UV Globular Cluster Survey (HUGS) \citep{Nardiello2018} and spectroscopic information from the literature.

\section{Asteroseismic analysis}

In a diagram of [$\alpha$/Fe] versus [Fe/H], the Milky Way (MW) disc-like components separate into two distinct sequences: one labeled $\alpha$-rich and the other $\alpha$-poor. The $\alpha$-rich component is known for its narrow age distribution, and consequently, its narrow mass distribution \citep[e.g.][]{Miglio2021}. From the catalogues of Willett et al., we selected giants belonging to the $\alpha$-rich population of the MW according to the selection criterion $[\alpha\rm/Fe] > 0.1 - 0.18\cdot[Fe/H]$, and split them into three [Fe/H] bins. The bin ranges were chosen to obtain similar numbers of HeCB and RGB stars in each bin for both the \textit{Kepler} and K2 samples. We used the asteroseismic scaling relation
\begin{eqnarray}
\label{eq:01}
\frac{M}{\mathrm{M}_\odot} & = & \left(\frac{\nu _{\mathrm{max}}}{f_{\nu _{\mathrm{max}}}\nu _{\mathrm{max,}\odot}}\right) \left(\frac{L}{\mathrm{L_{\odot}}}\right) \left(\frac{T_{\mathrm{eff}}}{\mathrm{T_{eff,\odot}}}\right)^{-7/2},
\end{eqnarray}
as well as the code PARAM \citep{daSilva2006,Rodrigues2017} to calculate masses and performed quality cuts as detailed in Appendix A. Here, $M$, $L$, $T_{\rm eff}$ and $\nu_{\rm max}$ are the mass, luminosity, effective temperature and the frequency of maximum power, respectively, see e.g. \citet{Brogaard2023}. $f_{\nu_{\rm max}}$ represents the potential deviation of $\nu_{\rm max}$ from the scaling relation relative to the Sun. We assumed $f_{\nu_{\rm max}}$=1, following \citet{Li2024}, who showed that potential deviations from that assumption are at or below the few percent level and therefore not significant for our findings.

The stars were separated into low-luminosity RGB and HeCB based on $\nu_{\rm max}$ and $T_{\rm eff}$; RGB stars are those with 40 < $\nu_{\rm max}/\rm \mu Hz < 200$ while HeCB stars are selected according to 20 < $\nu_{\rm max}/$$\rm \mu Hz < 35$ and $T_{\rm eff}$ greater than a limit depending on metallicity to avoid RGB stars of similar $\nu_{\rm max}$. We used this procedure to be consistent between the \textit{Kepler} and K2 samples while facing the issue that the quality and length of the K2 time series do not allow high-probability determination of the evolutionary state from mixed mode period spacings \citep{Bedding2011, Kallinger2012}. For the \textit{Kepler} sample we obtained identical mass-loss results when adopting the asteroseismic evolutionary states. In the appendix, we give details of the two samples to demonstrate the similarities of the stellar properties within each metallicity bin.

\begin{figure*}
   \centering
    \includegraphics[width=\hsize]{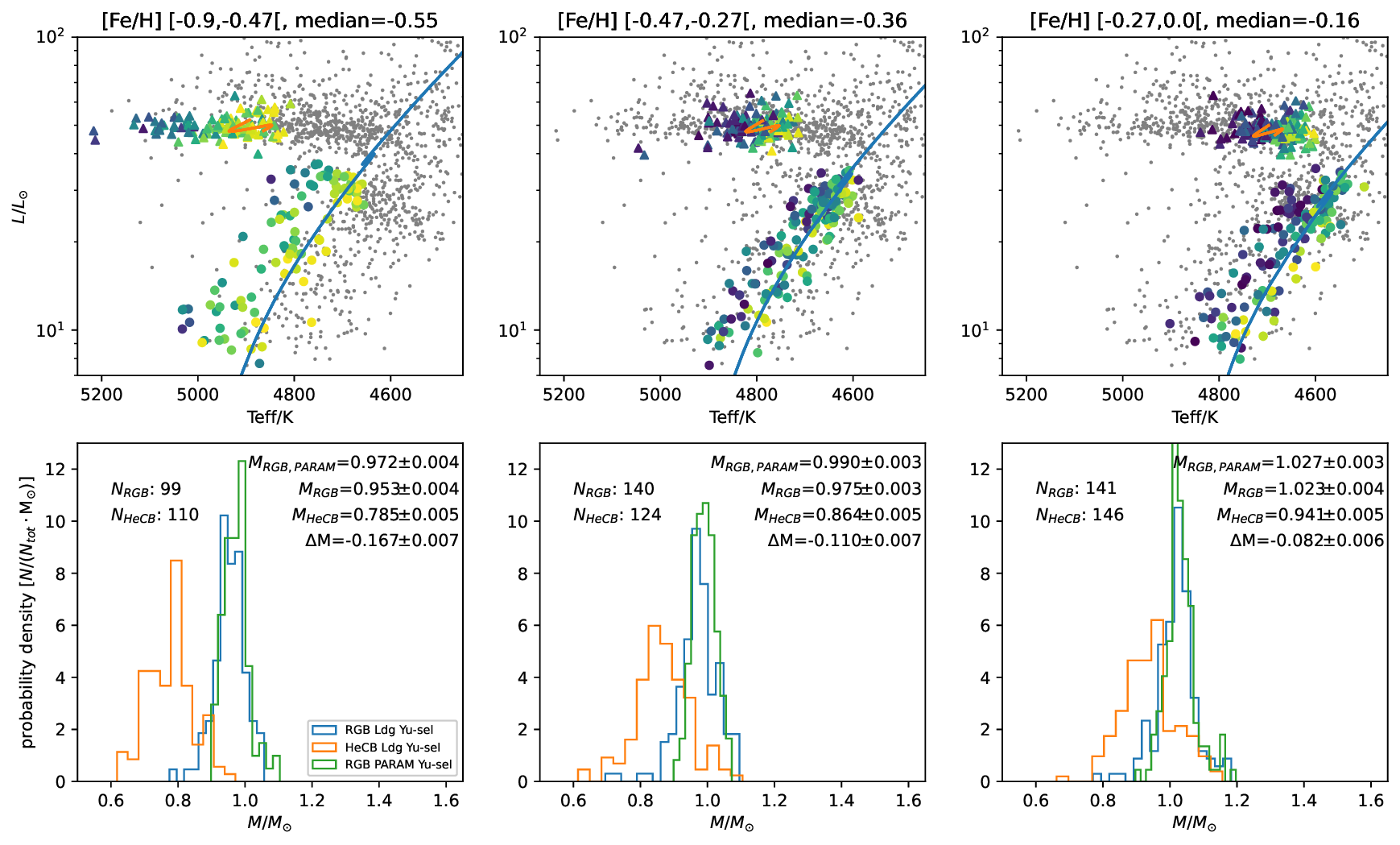}
      \caption{HR-diagrams and mass distributions of \textit{Kepler} high-$\alpha$ stars at three different metallicities. The [Fe/H] range and median in each panel is given at the top. 
      \textit{Top panels}: HR-diagrams of the selected stars. Colour-coding spans the metallicities in each bin with darker colours being more metal-poor. Solid lines represent MESA models of the median [Fe/H], [$\alpha$/Fe] and masses of each bin, shifted by -126 K, consistent with our analysis in Sect.~\ref{sec:HR_GC}.
      \textit{Bottom panels}: The \textit{Kepler} mass histograms based on asteroseismology from \citet{Yu2018} and Gaia DR3 parallax zero-points from \citet{Lindegren2021}. The mass distributions of RGB and HeCB stars are calculated using Eq.~\ref{eq:01}, and for the RGB stars also using PARAM with $\Delta\nu$ and $\nu_{\rm max}$ as input. The number of stars $N$ and the median mass of each evolutionary phase are shown along with the mass loss, $\Delta M$, the difference in median mass between the HeCB and RGB distributions.
      }
         \label{fig:K1HRmasses}
   \end{figure*}

Fig.~\ref{fig:K1HRmasses} shows the Hertzsprung-Russell (HR) diagrams of the selected \textit{Kepler} stars together with their mass distributions, median RGB and HeCB masses, and integrated mass loss $\Delta M$ as calculated from the difference between the RGB and HeCB mass medians for each metallicity bin. We performed Monte Carlo simulations to test whether the mass distributions of the RGB stars are consistent with a mono-mass distribution at each metallicity. The mass of each star in each bin was simulated by perturbing a single, average mass by errors drawn from the individual observed uncertainties. The recovered widths of the simulated distributions were fully compatible with the observed widths. This demonstrates that the observed mass distribution is dominated by observational errors, meaning that the true RGB mass distribution within each bin must be significantly narrower.

\begin{figure}
   \centering
    \includegraphics[width=\hsize]{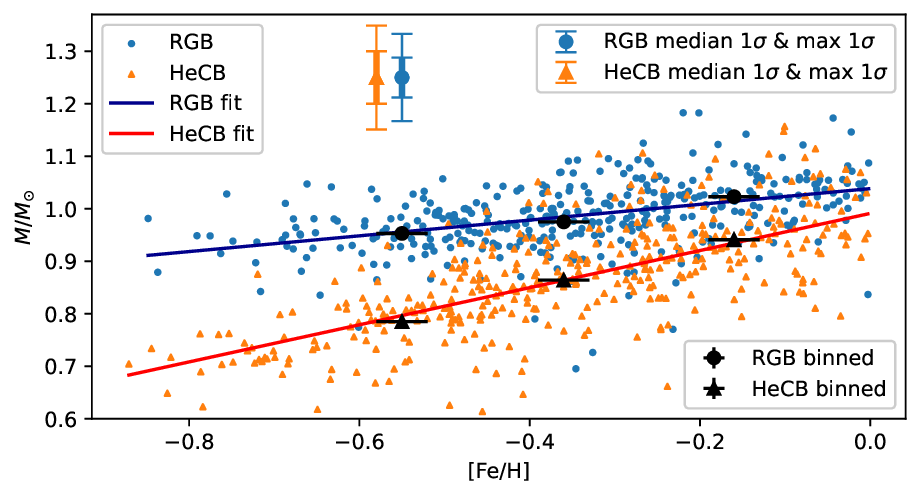}
      \caption{Mass vs [Fe/H] for RGB and HeCB \textit{Kepler} giants. Black binned data points correspond to the bins in Fig.~\ref{fig:K1HRmasses}. The lines represent linear fits to rolling medians. The blue and orange double-error-bars represent the median and maximum 1$\sigma$ mass uncertainty of the RGB and HeCB stars, respectively.}
         \label{fig:trend}
   \end{figure}

The average integrated mass loss is seen to increase with decreasing metallicity from 0.082 to 0.110 to 0.167 $\mathrm{M_{\odot}}$. This significant trend is also seen in Fig.~\ref{fig:trend} where we fitted linear relations to the full metallicity range and found clearly different slopes for RGB and HeCB stars, respectively. The scatter around the fit to the RGB stars is consistent with measurement uncertainties only, as seen from the sizes of the median 1$\sigma$ and maximum 1$\sigma$ error-bars of the individual stars.
From the difference between the lines of the RGB and HeCB stars, we obtained 
\begin{eqnarray}
\Delta M/\mathrm{M_{\odot}}=(-0.216\pm0.025) \times \rm [Fe/H] + (0.036\pm0.010). 
\end{eqnarray}

The trend is supported by our corresponding analysis of the K2 stars although shallower and at lower significance, see Appendix ~\ref{sec:K2}. We thus find clear evidence that RGB mass loss increases with decreasing metallicity. Since the RGB median mass is inversely correlated with metallicity, as expected for a nearly coeval population, the mass-loss trend could also be a trend with initial mass. More information is needed to establish whether metallicity or initial mass, or both, is the cause for the trend.

\section{HR and CMD diagram mass loss from GCs}

The RGB mass loss that we inferred from asteroseismology is independent of stellar models. Here, we examine its compatibility with model predictions for metal-rich GCs in HR and color-magnitude diagrams (CMDs). 

GCs are known to host multiple populations \citep{Gratton2012} where only the first generation (1G) is expected to have a Helium content similar to field stars \citep{Milone2018}. We therefore compare HeCB stars from our asteroseismic samples to 1G GC HeCB stars with similar [Fe/H] and [$\alpha$/Fe].

\subsection{HR comparison to metal-rich HeCB GC stars}
\label{sec:HR_GC}

In Fig.~\ref{fig:HR_GC} we compare the HR-diagram positions of HeCB stars from our \textit{Kepler} and K2 samples and two GCs with a metallicity close to [Fe/H]=$-0.50$ as detailed in the top panel legend. In particular, the HeCB stars in NGC\,6304 have [Fe/H], [$\alpha$/Fe] and $T_{\rm eff}$ measured by APOGEE DR17 \citep{VACSchiavon2024}, the same source we used for the \textit{Kepler} and K2 stars. This is therefore a direct $T_{\rm eff}$ comparison of HeCB stars in a GC and the field with the least possible amount of relative errors. Similarly, the chosen HeCB stars from NGC\,6352 sample well the colour-extent of the 1G HeCB stars in this cluster and have spectroscopic measurements \citep{Feltzing2009}, although not by APOGEE.
For both GCs, luminosities were derived using the spectroscopic $T_{\rm eff}$ and [Fe/H] values together with $K_S$ magnitudes \citep{Skrutskie2006}, reddening and true distance moduli as derived below from CMD fitting, and bolometric corrections \citep{Casagrande2014} for $K_S$. 

Comparing the GC stars directly to the field stars, their very similar effective temperatures suggest that their masses are also quite similar. Since there are much fewer GC stars and their individual uncertainties are larger, their mean $T_{\rm eff}$ values can be affected by sampling and evolution effects. However, as seen in Fig.~\ref{fig:NGC6352CMD}, the stars with spectroscopic $T_{\rm eff}$ measurements sample well the colour-extent of the 1G HeCB stars. Removing the most outlying star at either end changes the mean $T_{\rm eff}$ from 4967\,K by $-11$ or $+8$\,K, and the median is 4950\,K in all three cases. Resampling $T_{\rm eff}$ of nine stars randomly from the all 1G photometric members from Fig.~\ref{fig:NGC6352CMD} 1000 times results in a mean $T_{\rm eff}$ of 4985\,K with a maximum deviation of the median less than 20 K. We made similar evaluations for NGC\,6304 although less direct through the use of Gaia photometry, since the stars with spectroscopy does not have HST photometry.
Systematic errors in spectroscopic $T_{\rm eff}$ estimates can typically be at a level of about 80 K \citep{Bruntt2010}.  However, in our case, NGC\,6304 was spectroscopically analysed with the exact same instrument and method as the field stars, which should minimise systematics in the relative comparison.

The observed variation between models of different masses, as indicated by black double-headed arrows in the bottom panel of Fig.~\ref{fig:HR_GC} can be used to translate the observed $T_{\rm eff}$ differences into an estimate of a mass-difference. We used stellar model tracks computed using MESA v11701\footnote{MESA inlist is available on request.} \citep{Paxton2011, Paxton2013, Paxton2015, Paxton2018, Paxton2019} with inputs as described in \citep{Matteuzzi2023}.
The temperature scale of these models was found to be too high compared to the observations. There are several uncertainties in current stellar models, which can cause such offsets \citep[e.g., see][]{Silva2020}. One common issue is the adoption of mixing length theory \citep[e.g., see][]{Joyce2023} and the connected Solar calibration procedure to obtain the mixing length parameter, also connected to the choice of surface boundary condition \citep{Salaris2018}. To mitigate this to the best of our ability, all the model predictions in Fig.~\ref{fig:HR_GC} are shifted by -126 K so that the median $T_{\rm eff}$ of a 0.80$\mathrm{M_{\odot}}$ HeCB model track agrees with the median observed $T_{\rm eff}$ of the \textit{Kepler} stars, which also have a median mass very close to 0.80$\mathrm{M_{\odot}}$. While this procedure may not be accurate, it is encouraging that this $T_{\rm eff}$ shift also results in good agreement with the observed median luminosity of the \textit{Kepler} stars, which only improves if one considers that they have slightly higher than primordial Helium content (as the models in red colour assume). Additionally, the model tracks in Fig.~\ref{fig:K1HRmasses} show a good match to both RGB and HeCB stars in all three [Fe/H] bins with this same offset applied, even though it was determined at one specific metallicity.

The masses of the HeCB GC stars are found to be 0.03-0.07 $\mathrm{M_{\odot}}$ lower than the field stars when relying on the relative $T_{\rm eff}$-mass dependence of the shifted models. 
This is seen from the distances to the vertical coloured lines in the bottom panel of Fig.~\ref{fig:HR_GC}, which mark the mean $T_{\rm eff}$ values of the clusters corrected to the median metallicity of the field stars, $\rm [Fe/H]=-0.50$. Accordingly, our earlier considerations about potential uncertainties on the median $T_{\rm eff}$ GC values at a level below 20\,K suggest an uncertainty of less than 0.05 $\mathrm{M_{\odot}}$. 
We emphasise that this is a relative comparison between field and GC star masses, and that the models are only used to translate differences in $T_{\rm eff}$ and composition into differences in mass.

Comparing instead the median luminosities of the GC stars to that of the \textit{Kepler} stars in the top panel of Fig.~\ref{fig:HR_GC}, the HeCB median mass would be $-0.01$ to 0.04~$\mathrm{M_{\odot}}$ lower than that of the \textit{Kepler} stars, but this is much more uncertain, since there are few HeCB GC stars with spectroscopic measurements, and part of the luminosity difference could be due to evolution. If relying on the faintest CHeB star in each GC to avoid this, then the HeCB mass in the two clusters would instead be about 0.10 $\mathrm{M_{\odot}}$ lower than the \textit{Kepler} median, serving as a maximum difference estimate.
Since the GCs are likely older than the field stars, a mass difference is expected also for the RGB stars, which would result in very similar RGB mass loss for the metal-rich GCs and the \textit{Kepler} and K2 field stars. We investigate this in the following subsection.

\subsection{RGB mass loss from GC CMDs}

Fig.~\ref{fig:NGC6352CMD} shows the HST $F606W-F814W,F606W$ CMD of NGC\,6352 from the final data release \citep{Nardiello2018} of the HUGS survey \citep{Piotto2015} over-plotted with isochrones and zero-age horizontal branches (ZAHBs) from the Victoria models \citep{VandenBerg2014, VandenBerg2024}. 
HeCB stars were separated into 1G and 2G using the $F275W-F336W$ vs. $F336W-F438W$ diagram as in \citet{Tailo2022}.

\begin{figure}
   \centering
    \includegraphics[width=\hsize]{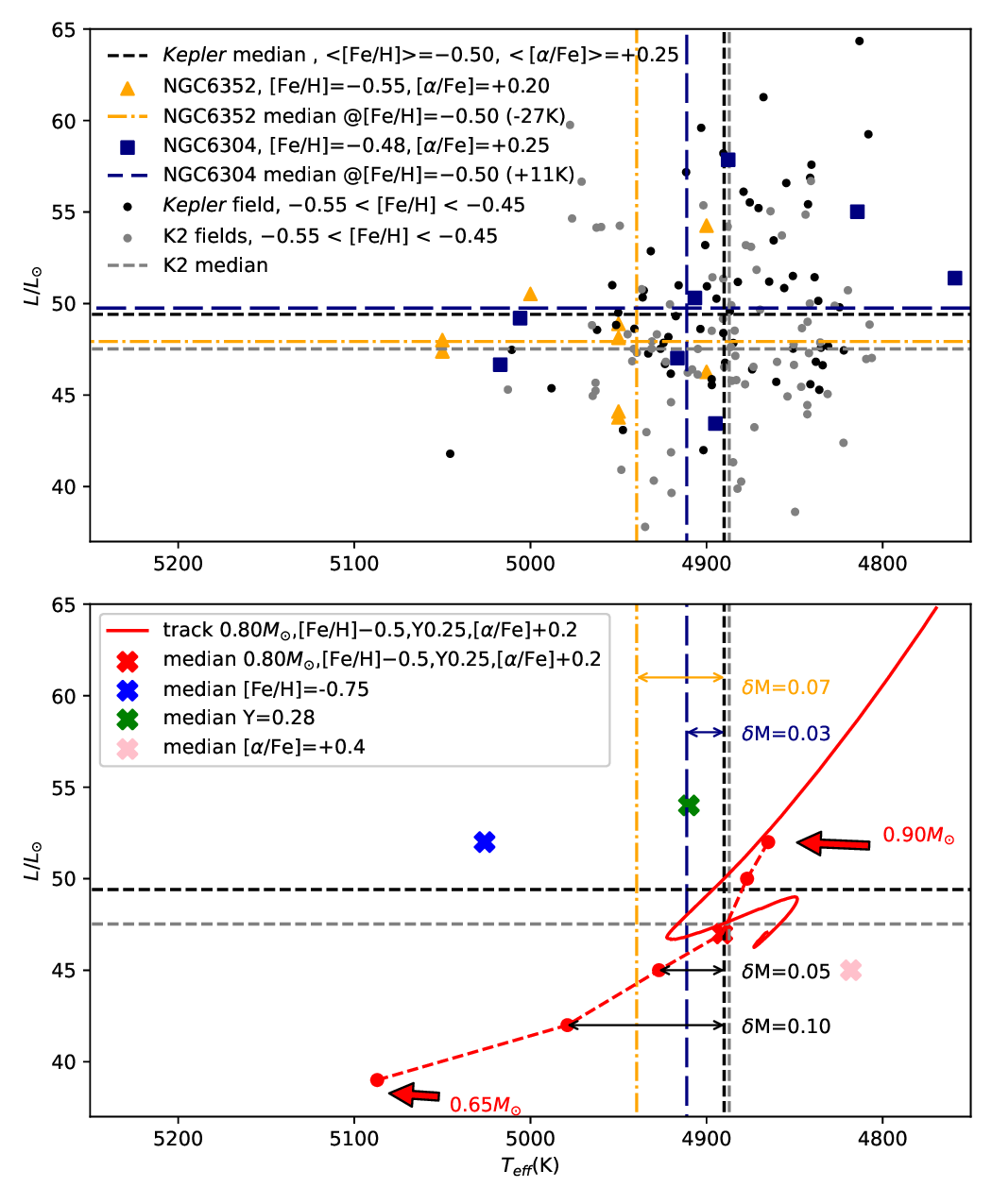}
      \caption{HR diagrams of $\alpha$-rich HeCB stars in the \textit{Kepler} and K2 fields compared to metal-rich GC HeCB stars and model predictions.
      \textit{Top panel}: \textit{Kepler} and K2 HeCB stars with $-0.55$ < [Fe/H] < $-0.45$ compared to GC stars of similar metallicity. The median colour of HeCB stars in each GC is corrected to $\mathrm{[Fe/H]=-0.50}$ based on model comparisons in the bottom panel.
      \textit{Bottom panel}: MESA HeCB stellar model track and median locations of various masses and compositions compared to observed median values. All model $T_{\rm eff}$ values are shifted by -126 K to reach agreement with the observed \textit{Kepler} median $T_{\rm eff}$ at the corresponding observed 0.8 $\rm M_{\odot}$. Crosses of different colours mark median locations of HeCB stars of different compositions. Model based estimates of mass-difference to temperature-difference are illustrated by black double-arrows. Coloured double-arrows mark the corresponding observed mass-differences between the \textit{Kepler} and GC HeCB medians. 
      }
         \label{fig:HR_GC}
   \end{figure}

\begin{figure}
   \centering
    \includegraphics[width=\hsize]{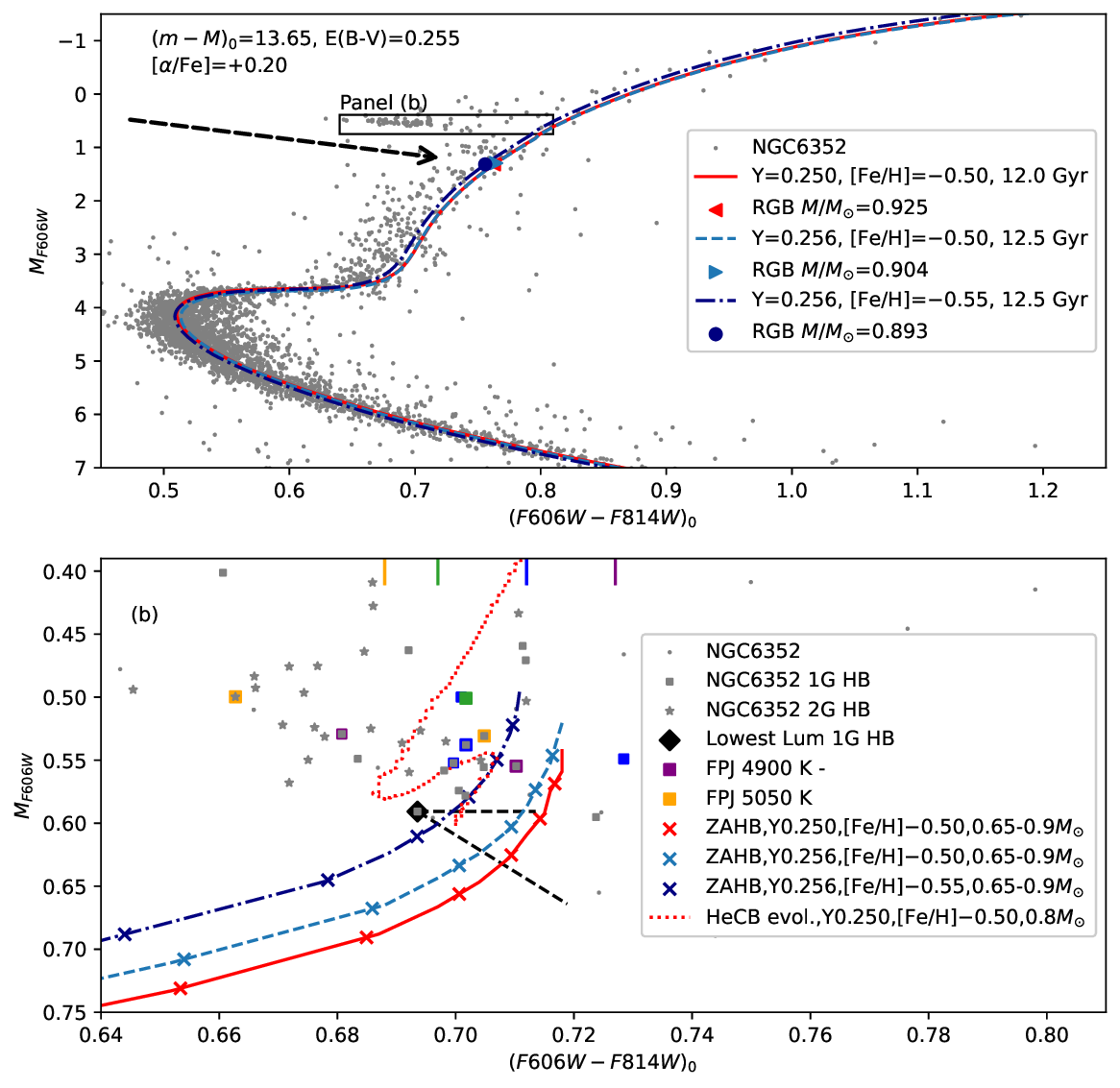}
    \caption{HST CMDs of NGC\,6352 with Victoria isochrones and ZAHBs and a MESA HeCB evolutionary track. \textit{Panel a:} Full CMD. Parameters equal for all models are given in the top left corner. The black dashed arrow is the reddening vector. \textit{Panel b:} Zoom on the RHB. 1G stars are marked with squares and 2G stars are marked with star symbols. Stars with spectroscopic measurements are marked with squares, purple corresponding to 4900\,K, blue is 4950\,K, green is 5000\,K, and yellow is 5050\,K. The $F606W-F814W$ colours of the corresponding temperatures are marked with vertical lines at the top border. The lowest luminosity RHB star is marked and dashed lines extending from it are used to estimate the HeCB mass at their intersections with ZAHBs. The sloped dashed line is along the reddening vector.
    }
         \label{fig:NGC6352CMD}
   \end{figure}

The models were matched to the observations employing the same nine HeCB GC stars with spectroscopic $T_{\rm eff}$ measurements used in Fig.~\ref{fig:HR_GC}; $E(B-V)$ was first adjusted to obtain the best average agreement between the spectroscopic $T_{\rm eff}$ values and corresponding photometric ones from $F606W-F814W$ and the calibration by \citet{Casagrande2014}. Then, the apparent distance modulus and age was found from a model-match to the main-sequence, turn-off and early SGB. 

We inferred the minimum HeCB mass by comparing the absolute magnitude of the least luminous 1G HeCB star (black diamond in panel b of Fig.~\ref{fig:NGC6352CMD}) to ZAHB loci.
This can be accomplished along the sloped dashed line, which represents the direction of both the reddening vector and a change in metallicity. Errors in metallicity, reddening, and/or differential reddening therefore shift stars or the ZAHBs along this line. However, because of potential systematic uncertainties caused by e.g. mixing length theory and surface boundary conditions, it is also likely that the ZAHB should be shifted horizontally. Such a shift is unlikely to be past the red ZAHB shown, unless there are errors in the relative temperature scale, since the corresponding isochrone is already at the cool red edge of the observed RGB (see panel (a) of Fig.~\ref{fig:NGC6352CMD}). 
Thus, one either obtains a minimum HeCB mass of about 0.86, 0.82 or 0.78~$\mathrm{M_{\odot}}$ depending on the assumed composition from a horizontal match or lower values of about 0.79, 0.75 or 0.72 $\mathrm{M_{\odot}}$ from a matches along the reddening line.
These HeCB mass estimates were subtracted from the corresponding low-luminosity RGB masses from the CMD-matched model close to the RGB bump in panel (a) of Fig.~\ref{fig:NGC6352CMD} to obtain the mass-loss predictions. This yielded estimates of the integrated RGB mass loss in the ranges 0.07-0.14, 0.08-0.15, and 0.11-0.17\,$\mathrm{M_{\odot}}$ for the chosen compositions. As seen, the range is not strongly dependent on composition, because changes to the RGB mass partly compensates the changes to the HeCB mass estimates. We adopt the range 0.11-0.17 $\mathrm{M_{\odot}}$ corresponding to the metallicity [Fe/H]=$-0.55$ measured by \citet{Feltzing2009} and with a helium content of $Y=0.256$. Since we cannot know which end of the range to prefer, we take as our best estimate the middle value and distance to the range ends as the uncertainty, to obtain $\Delta M=0.14\pm0.03 \mathrm{M_{\odot}}$. In the appendix, we repeat the procedure for NGC\,6304 for which we obtain $\Delta M=0.13\pm0.04 \mathrm{M_{\odot}}$. Thus, the mass loss in these GCs is very similar to that inferred from asteroseismology of field stars at similar metallicity.

\section{Discussion, conclusions, and outlook}
\label{sec:conclusions}

\begin{figure}
   \centering
    \includegraphics[width=\hsize]{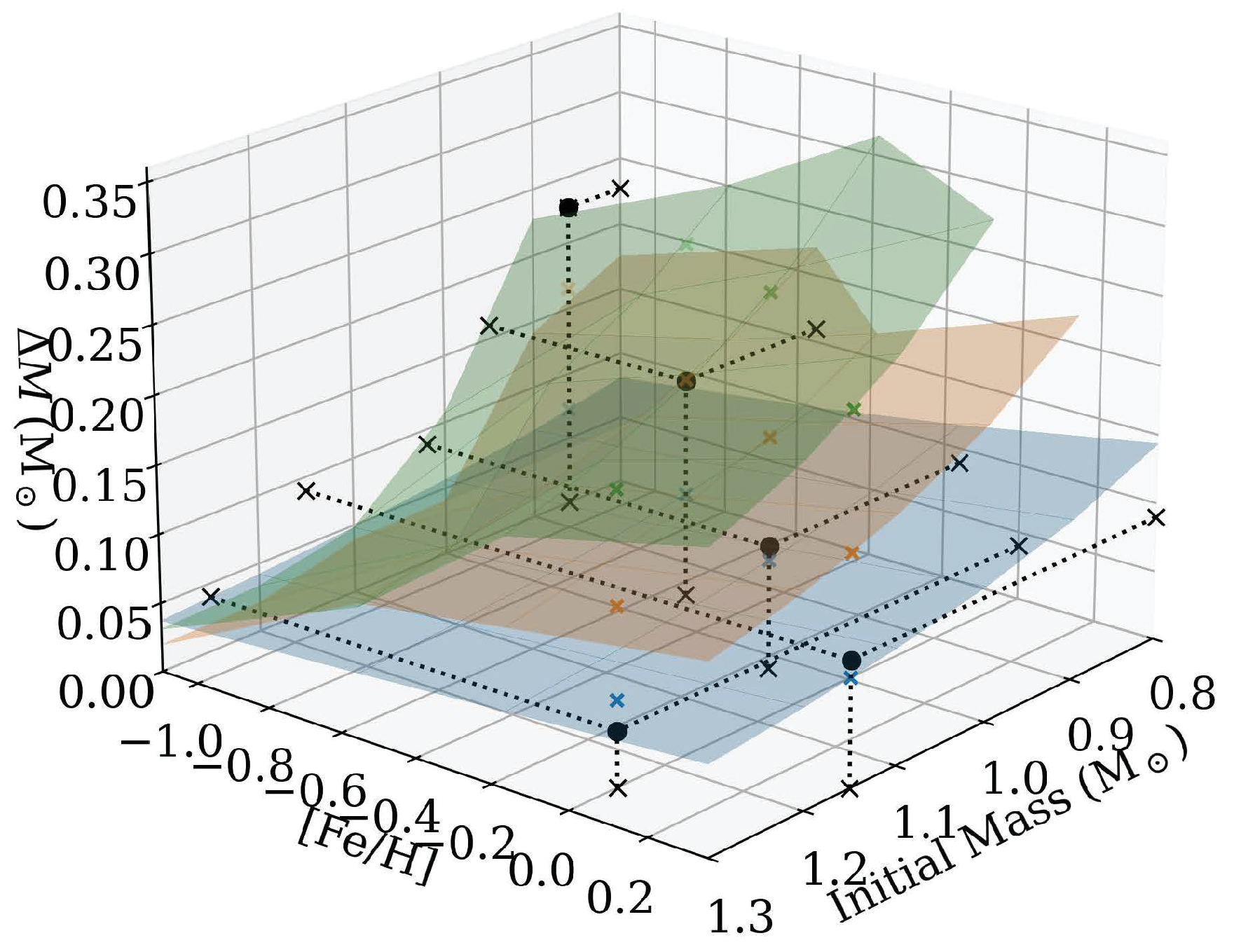}
      \caption{Comparison of integrated mass-loss estimates from the GC M4 \citep{Tailo2022}, our most metal-poor and most metal-rich bins of $\alpha$-rich field stars, and the open clusters M67 \citep{Stello2016} and NGC\,6791 \citep{Miglio2012} to predictions from Reimers mass-loss law for different choices of $\eta$. Blue: $\eta$=0.2, orange: $\eta$=0.4, green: $\eta$=0.6. 
      }
         \label{fig:reimers}
   \end{figure}

We provide novel empirical constraints on the integrated RGB mass loss, crucially closing the gap between GCs, field stars and open clusters with asteroseismic constraints.
\begin{itemize}
    \item By comparing the median mass of RGB and HeCB stars in nearly mono-age mono-metallicity populations of giants in the MW high-[$\alpha$/Fe] sequence, we find $\Delta M$ to increase with decreasing metallicity/mass.
    \item Such an increase in $\Delta M$ extends to M4, where constraints from asteroseismololgy and CMDs agree \citep{Tailo2022}.
    \item GCs were reported in the literature to have increasing $\Delta M$ \citep{Tailo2020} towards higher metallicity. Our RGB mass-loss estimates for NGC\,6352 and NGC\,6304 ([Fe/H]$\sim$ -0.5) are significantly smaller than in \citealt{Tailo2020} and in agreement with our asteroseismic findings for \textit{Kepler} and K2 field stars at similar metallicities and masses, suggesting that the mechanism and efficiency of RGB mass loss is the same for field and GC stars.
    \item Current RGB mass-loss laws are unable to match all the asteroseismic integrated mass-loss estimates using a single value for the free parameter $\eta$. Fig.~\ref{fig:reimers} demonstrates this for Reimers' law in comparison to our estimates from $\alpha$-rich field stars in this work, as well as the GC M4 and the open clusters M67 and NGC\,6791. We found comparable issues for similar prescriptions (see also Fig. 4 in  \citealt{Catelan2009}).

\end{itemize}
The situation is further complicated by the fact that RGB mass-loss estimates in GCs suggest an increase in mass loss with increasing metallicity/mass while our present findings suggest a decrease with increasing metallicity/mass.
It might therefore be that the true mass-loss prescription is first increasing with increasing mass/metallicity at low metallicities and after some specific threshold the trend reverses. Further theoretical developments regarding the mechanism responsible for the mass loss process on the RGB, combined with more precise asteroseismic measurements for additional GCs — which could be provided by a space mission like Haydn \citep{Miglio2021b} — pave the way for transforming our understanding of mass loss from a rudimentary calibration to a well-defined physical insight.

\begin{acknowledgements}

We thank Don VandenBerg for providing Victoria isochrones and ZAHBs.

This work has made use of data from the European Space Agency (ESA) mission {\textit{Gaia}} (\url{https://www.cosmos.esa.int/Gaia}), processed by the {\textit{Gaia}} Data Processing and Analysis Consortium (DPAC,\url{https://www.cosmos.esa.int/web/Gaia/dpac/consortium}). Funding for the DPAC has been provided by national institutions, in particular the institutions participating in the {\textit{Gaia}} Multilateral Agreement.\\
This paper includes data collected by the Kepler mission. Funding for the Kepler mission is provided by the NASA Science Mission directorate. Some of the data presented in this paper were obtained from the Mikulski Archive for Space Telescopes (MAST). STScI is operated by the Association of Universities for Research in Astronomy, Inc., under NASA contract NAS5-26555. Support for MAST for non-HST data is provided by the
NASA Office of Space Science via grant NNX09AF08G and by other grants and contracts.\\
Funding for the Stellar Astrophysics Centre was provided by The Danish National Research Foundation (Grant agreement no.: DNRF106).\\
AM, EW, KB and WEvR acknowledge support from the ERC Consolidator Grant funding scheme (project ASTEROCHRONOMETRY, \url{https://www.asterochronometry.eu}, G.A. n. 772293).

\end{acknowledgements}

\bibliographystyle{aa} 
\bibliography{References-1} 

\begin{appendix}

\section{Asteroseismic measurement details}    

In this section we provide details on our method to derive asteroseismic integrated mass-loss estimates.

\begin{figure*}[h!]
   \centering
    \includegraphics[width=\hsize]{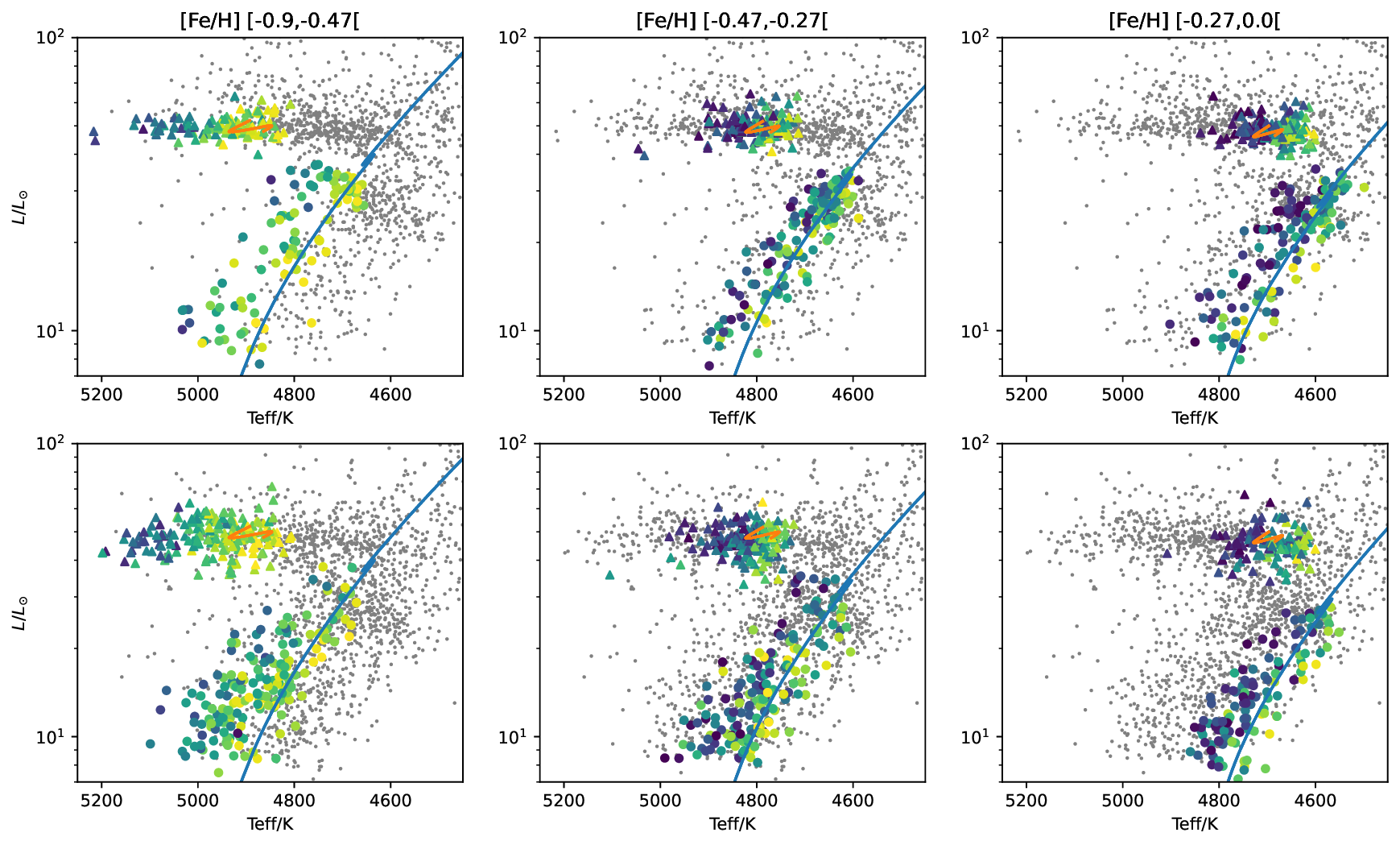}
      \caption{HR diagrams within the chosen metallicity bins for the \textit{Kepler} and K2 samples with the quality cuts applied.
      }
         \label{fig:hr}
   \end{figure*}

\begin{figure*}[h!]
   \centering
    \includegraphics[width=\hsize]{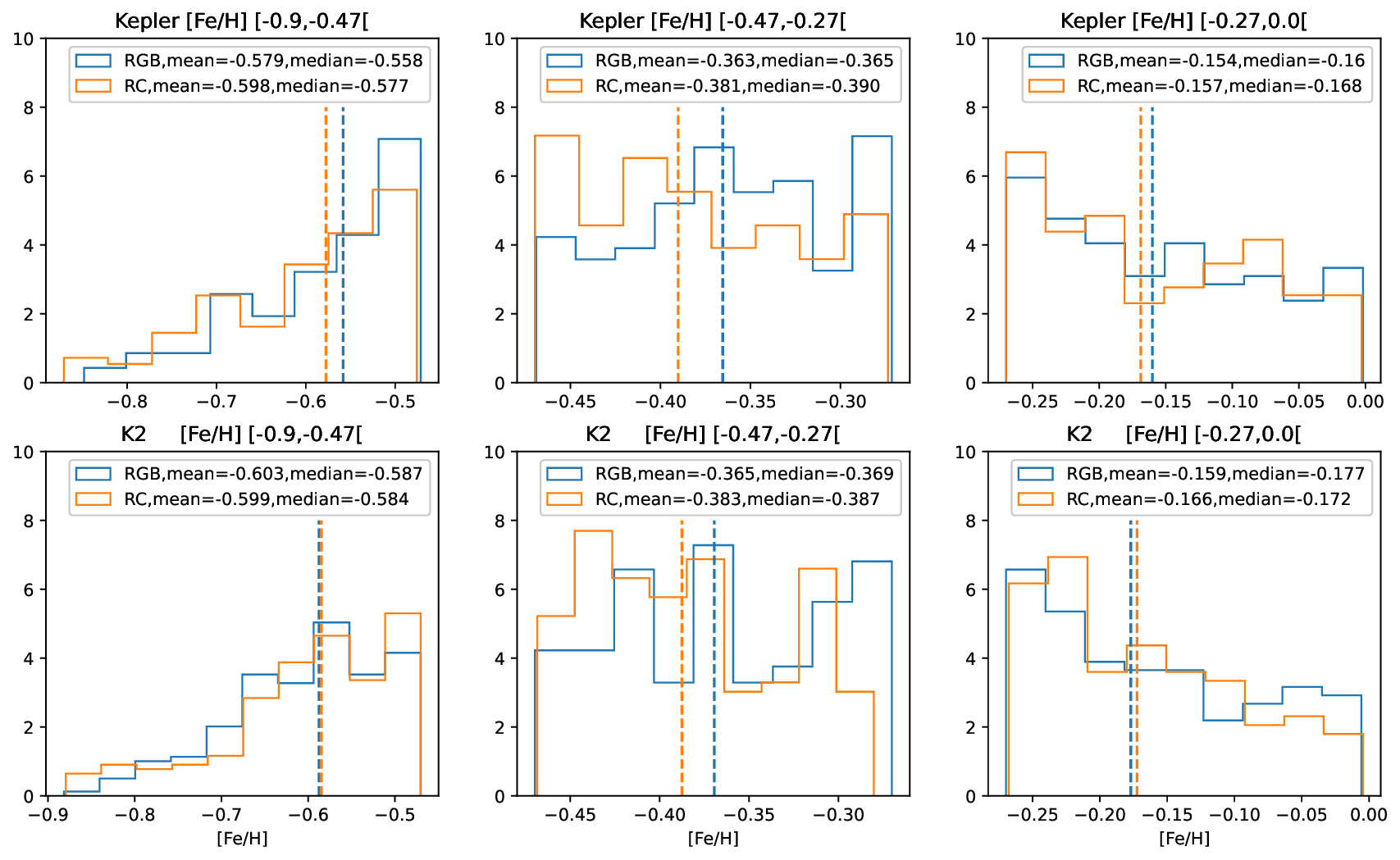}
      \caption{Histograms of metallicity distributions within the chosen metallicity bins for the \textit{Kepler} and K2 samples with the quality cuts applied. Vertical dashed lines correspond to the median [Fe/H] of RGB and HeCB stars, respectively.
      }
         \label{fig:feh}
   \end{figure*}

\begin{figure*}[h!]
   \centering
    \includegraphics[width=\hsize]{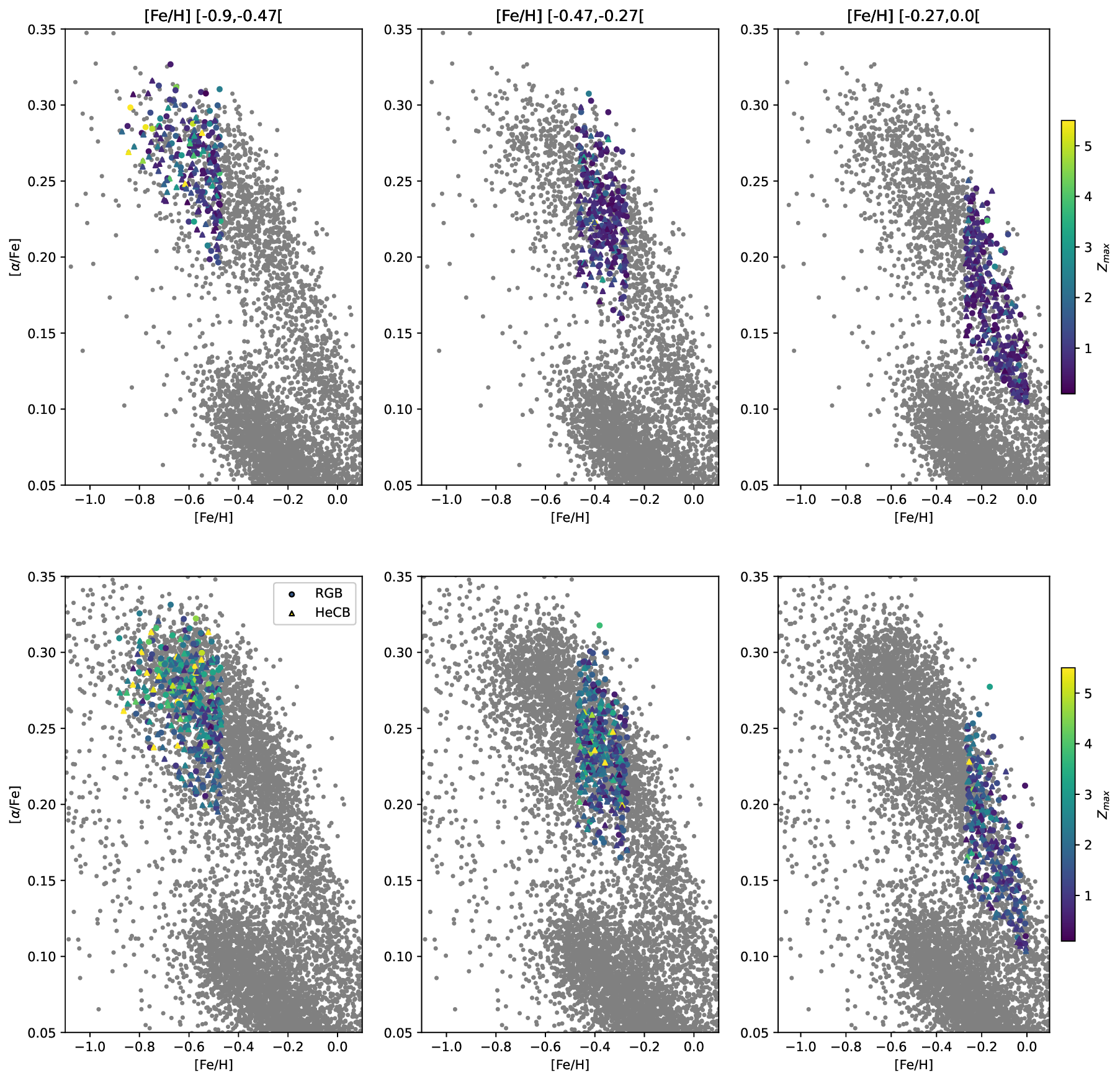}
      \caption{[$\alpha$/Fe] vs [Fe/H] distributions of the selected stars. Grey dots are the full samples. Coloured circles and triangles indicate the selected RGB and HeCB stars. Colour-coating is according to $Z_{\rm max}$, the distance from the MW disc plane.Top row are for \textit{Kepler} stars, bottom row the K2 stars. 
      }
         \label{fig:alpha-feh}
   \end{figure*}

As mentioned in the main text, we selected giants belonging to the $\alpha$-rich population of the MW according to the selection criterion $[\alpha\rm/Fe] > 0.1 - 0.18\cdot[Fe/H]$, and split them into three [Fe/H] bins (see Figs.~\ref{fig:hr}--~\ref{fig:alpha-feh}). The bin ranges were chosen to obtain similar numbers of HeCB and RGB stars in each of the three bins for both the \textit{Kepler} and K2 samples. RGB and HeCB stars were separated using different limits on $\nu_{\rm max}$ and $T_{\rm eff}$ as detailed in Table ~\ref{tab:quality}.  

We then applied quality cuts as given in Table ~\ref{tab:quality}. We removed stars with a $\nu_{\rm max}$ uncertainty above 5\% and stars with an RGB mass from PARAM with an uncertainty above 7\%. Constraints were also applied for the parallax and parallax uncertainty as explained below in Section~\ref{sec:K2}. The limits were chosen as a compromise between precision and the number of stars remaining in the bins. Additionally, cuts were made in the HR diagram to remove stars that appeared significantly hotter than the general RGB and fainter than the HeCB stars, since these are either stars whose parameters have been altered through binary evolution, or stars with bad measurements. Despite these cuts, few such over-massive stars remain in the samples, and we therefore also cut away stars more massive than the 95th percentile for each population separately. However, this did not alter significantly the values of the median masses or the integrated RGB mass loss.

\begin{table}[hbt!]
\caption{Cuts \& constraints for selection of stars}  
\label{tab:quality}      
\centering                          
\begin{tabular}{l | c }
\hline\hline                 
Parameter  & constraint\\
\hline                                   
$[\alpha/\rm Fe]$ & $0.10-0.18\cdot$[Fe/H] \\
$\nu_{\rm max}(\mu$Hz) & 20-200 \\
$M(\mathrm{M_{\odot}})$ & 0.45-1.55 \\
$\sigma_{\nu_{\rm max}}$ (\%) & < 5 \\
Parallax (mas) & > 0.34 \\
$\sigma_{\rm Parallax}$ (\%) &  < 5 \\
HR-diagram & outliers removed\\
Mass distribution & cut above 95\% percentile\\
\hline
RGB only: & \\
$\nu_{\rm max}(\mu$Hz) & > 40 \\
$\sigma_{M/M_{\odot,\rm PARAM}}$ (\%)& < 7 \\
\hline
HeCB only: & \\
$\nu_{\rm max}(\mu$Hz) & < 35  \\
$T_{\rm eff}$ (K) & [Fe/H] dependent low limit \\
\hline                                   
\end{tabular}
\end{table}

The HR diagrams of our selected stars are shown in Fig.~\ref{fig:hr}. The $T_{\rm eff}$ distributions of the RGB and HeCB stars cover similar ranges, minimising the possibility of the derived mass loss to be an artefact of systematics in the observed $T_{\rm eff}$ scale, since the mass of RGB and HeCB would be affected in the same way. Although not shown, we also checked that there is no correlation between mass and luminosity for the RGB stars (but there is a correlation for HeCB which is expected from stellar evolution). Therefore, it is unlikely that systematic errors in luminosity between RGB and HeCB stars are biassing our mass-loss estimates. 

Fig.~\ref{fig:feh} shows histograms of the [Fe/H] distributions of RGB and HeCB stars in each metallicity bin to demonstrate that they are very similar. Differences in [Fe/H] medians and means between \textit{Kepler}-K2 and RGB-HeCB are typically at or below 0.02 dex, sometimes as high as 0.03 dex. The general mass-change between panels suggest that a change of 0.01 dex in metallicity corresponds to a mass change of about 0.001 $\mathrm{M_{\odot}}$. Therefore, these small differences in metallicity have very limited influence on the derived masses and mass loss at the level of 0.002 $\mathrm{M_{\odot}}$.

For the \textit{Kepler} sample, measurement uncertainties on $\nu_{\rm max}$, $T_{\rm eff}$ and luminosity are in general small enough that the inferences using the full sample did not change significantly even if various quality-cuts were tried. For the final results we decided for consistency reasons to anyway apply the same quality cuts as for the analysis of the K2 sample.

\subsection{Kepler}

Figure~\ref{fig:K1mass} shows histograms of masses in the three metallicity-bins for RGB and HeCB stars, and values are given for the median mass and number of stars of each population in each bin along with $\Delta M$, the integrated RGB mass loss, calculated as the difference between the median mass of RGB and HeCB stars, respectively. Uncertainties on the numbers are calculated by assuming Gaussian uncertainties and adding random uncertainties to each observable of each star before calculating the median mass 1000 times in a Monte Carlo run.
For the RGB stars, there are two mass estimates; one based on $\nu_{\rm max}$ and luminosity using the Gaia parallax (Eq. 1), and another distance-independent estimate based on using both $\nu_{\rm max}$ and $\Delta\nu$ in the Bayesian tool PARAM. The two mass estimates are close but do not agree within mutual 1$\sigma$ uncertainties in two out of the three metallicity bins, reflecting that either uncertainties are slightly underestimated and/or are systematic in nature at this level. Among possible causes are small shifts to the parallax zero-points and/or the temperature scale. For stars in the RC we preferred not to use results from PARAM as they are found to be strongly dependent on the evolutionary tracks. The latter, by definition, occupy a rather small volume in the observables (L, $T_{\rm eff}$, $\nu_{\rm max}$, ...) and any systematic difference between models and observations is found to have a strong impact on the mass estimates, which we would rather avoid in our study.
In Fig.~\ref{fig:K1mass} each set of three horizontal panels  represent a different case. The top panels show the full sample of \textit{Kepler} high-$\alpha$ stars without quality cuts and adopting asteroseismic parameters from \citet{Yu2018}. The middle panel shows that no significant change occurs for the \textit{Kepler} sample due to applying our quality cuts. The bottom panel adopts alternative asteroseismic measurements provided by Y. Elsworth (private comm.) calculated following \citet{Elsworth2020}. This allows a more direct comparison to the K2 sample, which uses seismic values derived using that method.

\subsection{K2}
\label{sec:K2}
\begin{figure*}[h!]
   \centering
    \includegraphics[width=\hsize]{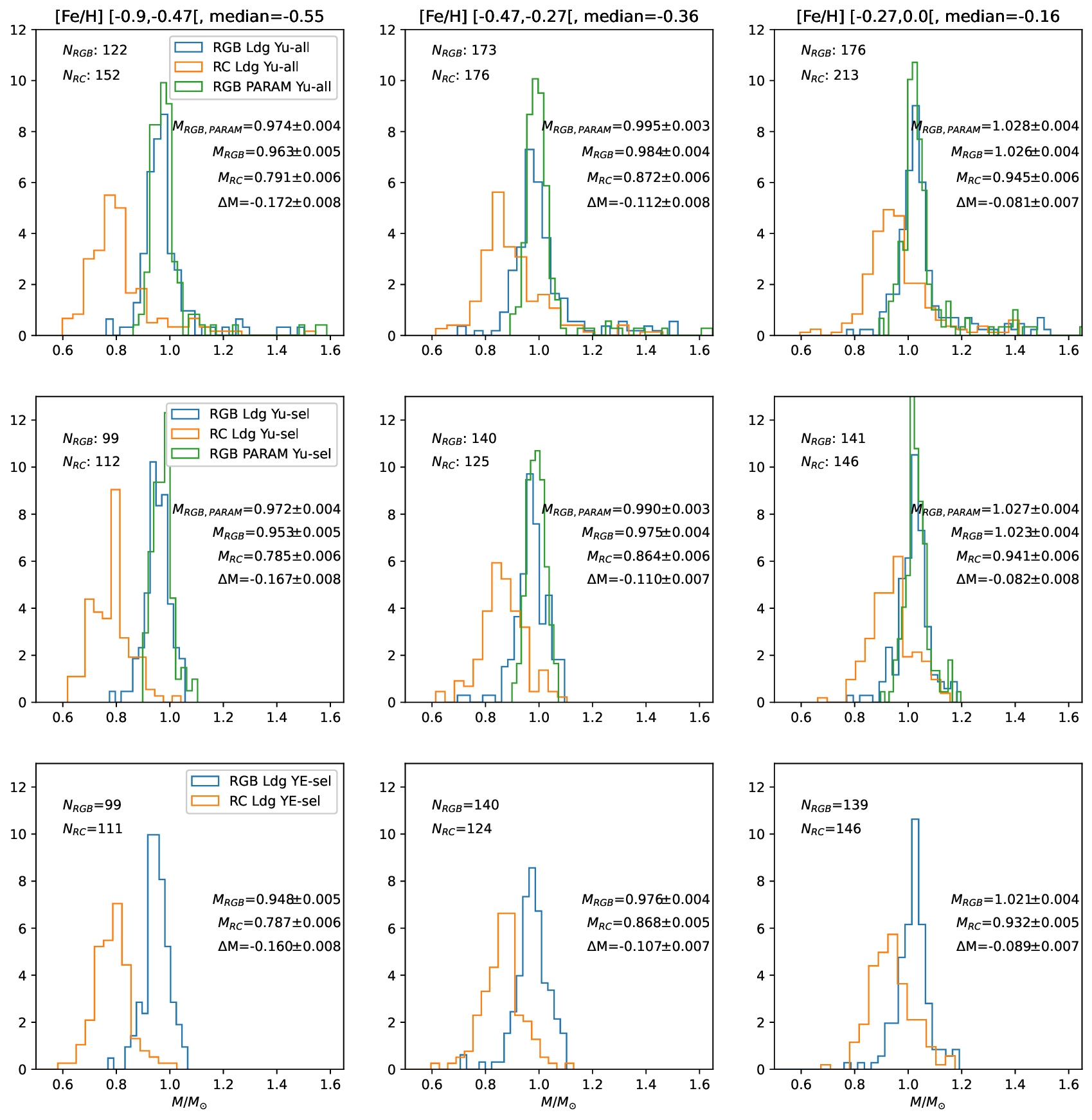}
      \caption{Mass distributions of stars in sub-samples of \textit{Kepler} high-$\alpha$ stars at three different metallicities. The metallicity range and median metallicity in each panel is given above the top row. Each panel shows the mass distributions or RGB and RC stars separately, calculated using the scaling relation with $\nu_{\rm max}$ and luminosity, and for the RGB stars also the mass from PARAM with $\Delta\nu$ and $\nu_{\rm max}$ as input. The number of stars  the median mass values of each evolutionary phase are given. Also stated is the mass loss given as $\Delta M$, the difference in mass between the RC and RGB phases.
      \textit{top panels}: All stars in the sample.
      \textit{middle panels}: With quality-cuts applied as described in the text. 
      \textit{bottom panels}: As middle panels, but using the average asteroseismic parameters from \citet{Elsworth2020} instead of \citet{Yu2018}.    
      }
         \label{fig:K1mass}
   \end{figure*}

In Fig.~\ref{fig:K1mass} we repeated the procedure for K2. Masses calculated from K2 data have significantly larger uncertainties because the time-series are much shorter and because the Gaia parallax zero-points are much more uncertain for the K2 fields than for the single \textit{Kepler} field, even in a relative sense from one K2 field to another.

In the first two rows we employed a fixed -17 $\mu$as parallax zero-point correction with, and without, quality cuts applied, respectively. With this zero-point we found reasonable agreement between the RGB masses calculated from $\nu_{\rm max}$ plus luminosity from parallax and from PARAM using $\nu_{\rm max}$ and $\Delta\nu$. This was not the case when using the \citet{Lindegren2021} correction for the K2 fields, as can be seen in the fourth row. Note however that the same mass-loss trend remains despite this issue with the absolute masses.

\begin{figure*}[h!]
   \centering
    \includegraphics[width=\hsize]{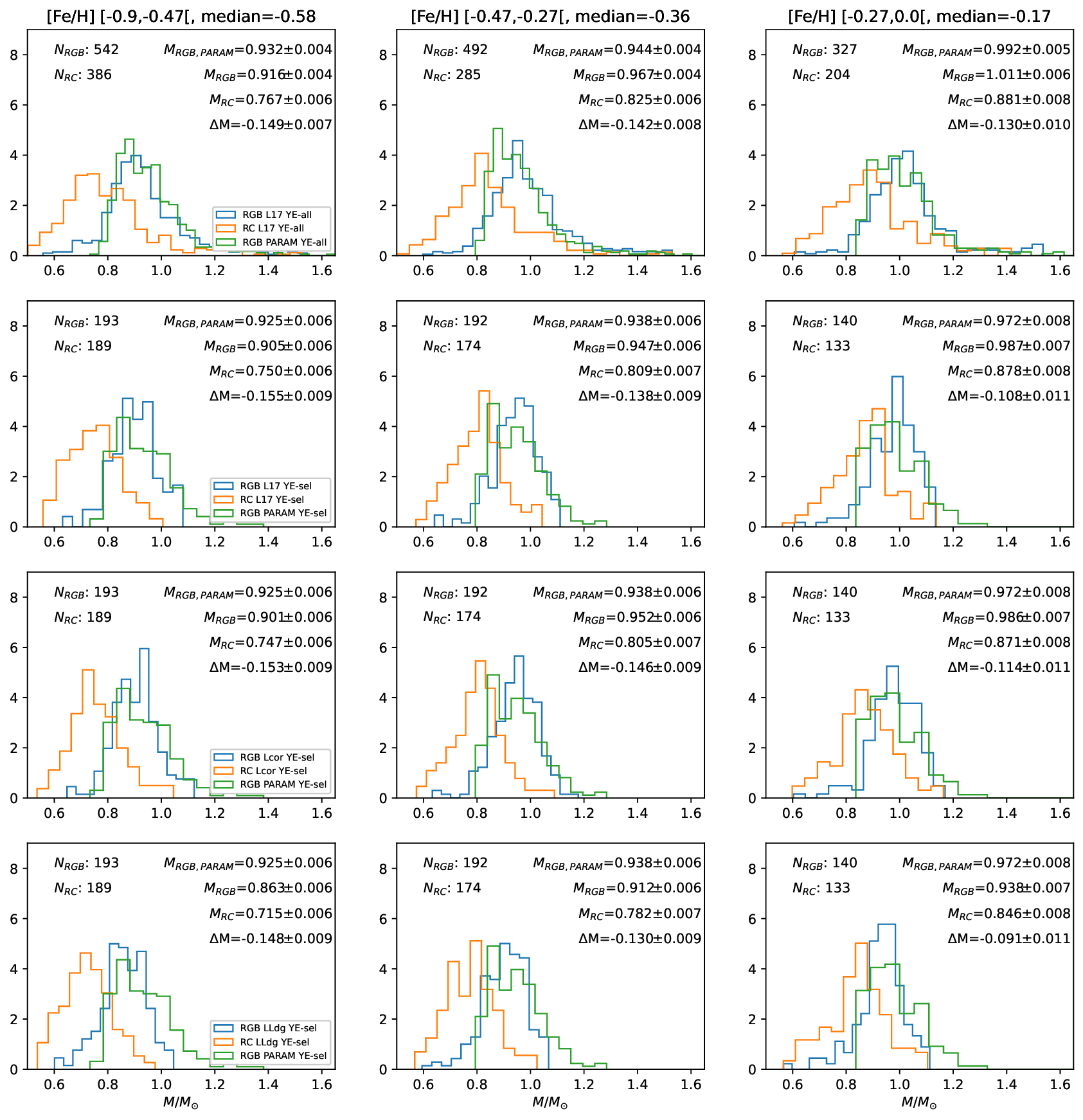}
      \caption{Similar to Fig.~\ref{fig:K1mass} but for the K2 sample. \textit{1. row}: All stars in the sample.
      \textit{2. row}: With quality-cuts applied as described in the text and using a constant -17 $\mu$as parallax zero-point offset. 
      \textit{3. row}: As 2. row, but using sector-based parallax zero-point offsets from \citet{Khan2023}.
      \textit{4. row}: As 2. row, but using parallax zero-point offsets from \citet{Lindegren2021}.
      }
         \label{fig:K2mass}
   \end{figure*}

To reduce potential problems from parallax zero-point errors we removed stars with $\tilde{\omega} < 0.34$ and $\sigma_{\tilde{\omega}}/\tilde{\omega} < 0.05$ to avoid cases with large errors (>5\%) on parallax, either systematic or statistical. All but the top row panels of Fig.~\ref{fig:K2mass} use these criteria.

Using a fixed parallax zero-point for all K2 fields is not necessarily correct, and \citep{Khan2023} used a sample of 7024 K2 stars with asteroseismic and parallax information to infer zero-points for each field. We repeated our procedure while implementing these field-by-field parallax zero-points, which resulted in median mass changes at or below the $1-\sigma$ level. These results are shown in the third row of panels in Fig.~\ref{fig:K2mass}.

The trend of the integrated mass loss for the K2 giants is similar to and consistent with that found for the \textit{Kepler} giants, although the slope is shallower. However, as seen by comparing numbers in Fig.~\ref{fig:K1mass} and Fig.~\ref{fig:K2mass}, the uncertainties are also larger, and the increase in mass loss between the metallicity bins depends on assumptions. To reach full $1\sigma$ agreement with the \textit{Kepler} results, an uncertainty on the median masses larger than the statistical one needs to be adopted, at the level of 0.02 $\mathrm{M_{\odot}}$ for K2 and 0.01 $\mathrm{M_{\odot}}$ for \textit{Kepler}. Thus, while the K2 result does support the \textit{Kepler} result, it is less significant on its own.

\FloatBarrier

\FloatBarrier

\FloatBarrier

\FloatBarrier

\FloatBarrier   

\section{RGB mass-loss dispersion}

\label{sec:sigma_deltaM}
\begin{figure*}[t]
   \centering
    \includegraphics[width=18cm]{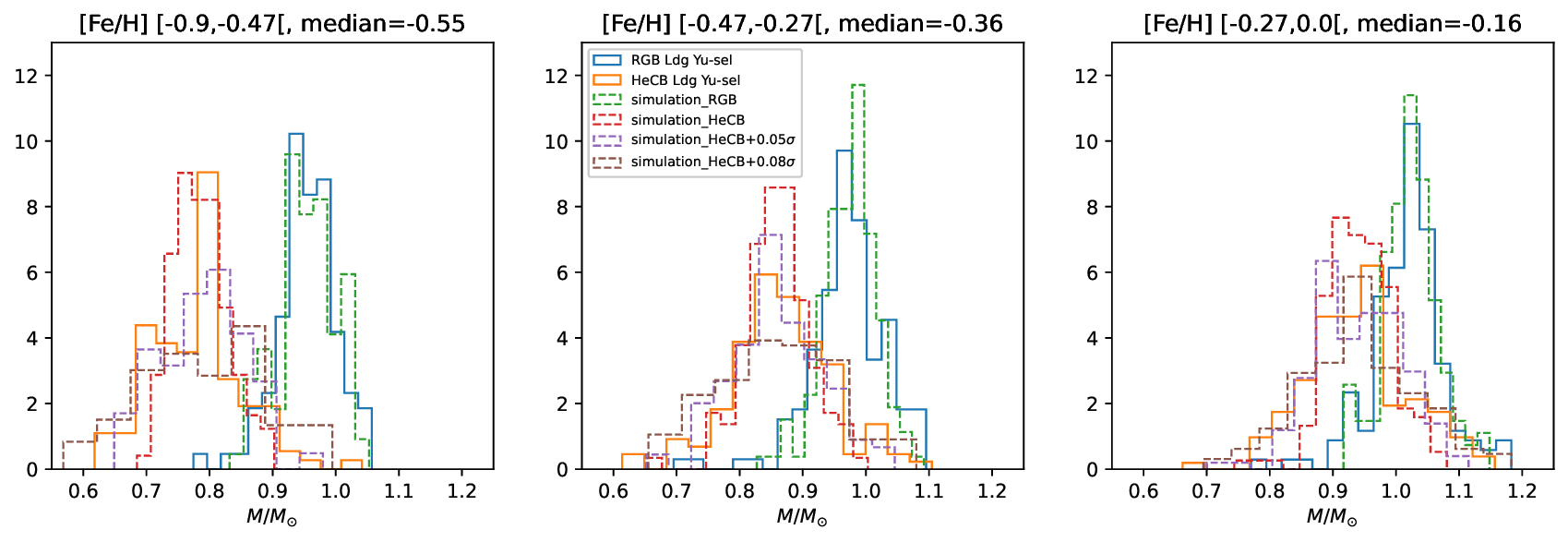}
      \caption{RGB mass-loss dispersion from observed \textit{Kepler} vs. simulated mass histograms.}
         \label{fig:dispersion}
   \end{figure*}

Fig.~\ref{fig:dispersion} shows the observed \textit{Kepler} mass distributions compared to simulations assuming the median mass represents stars with a single common mass at each [Fe/H] with the observed uncertainties from the real data. As seen, the distribution widths are compatible between observations and simulations for the RGB stars. For the HeCB stars, the simulations show narrower distributions than the observations unless additional scatter is applied. This additional scatter could arise due to underestimated uncertainties, but it could also be real. In the latter case, it could be interpreted as mass-loss dispersion since it is not present for the RGB stars. Two simulations are shown assuming a Gaussian mass-loss dispersion with $\sigma$ of 0.05 and 0.08 $\mathrm{M_{\odot}}$, respectively. Due to the asymmetry of the observed distributions, it is difficult to interpret how to get a best match. A by-eye comparison of distribution widths suggests that 0.05$\sigma$ is close to an upper limit and that 0.08 $\mathrm{M_{\odot}}$ is clearly too high.

\FloatBarrier 

\section{GCs}

In the main paper we used the CMD of the GC NGC\,6352 to infer the integrated RGB mass loss by comparing to models. 

\begin{figure}[h!]
   \centering
    \includegraphics[width=\hsize]{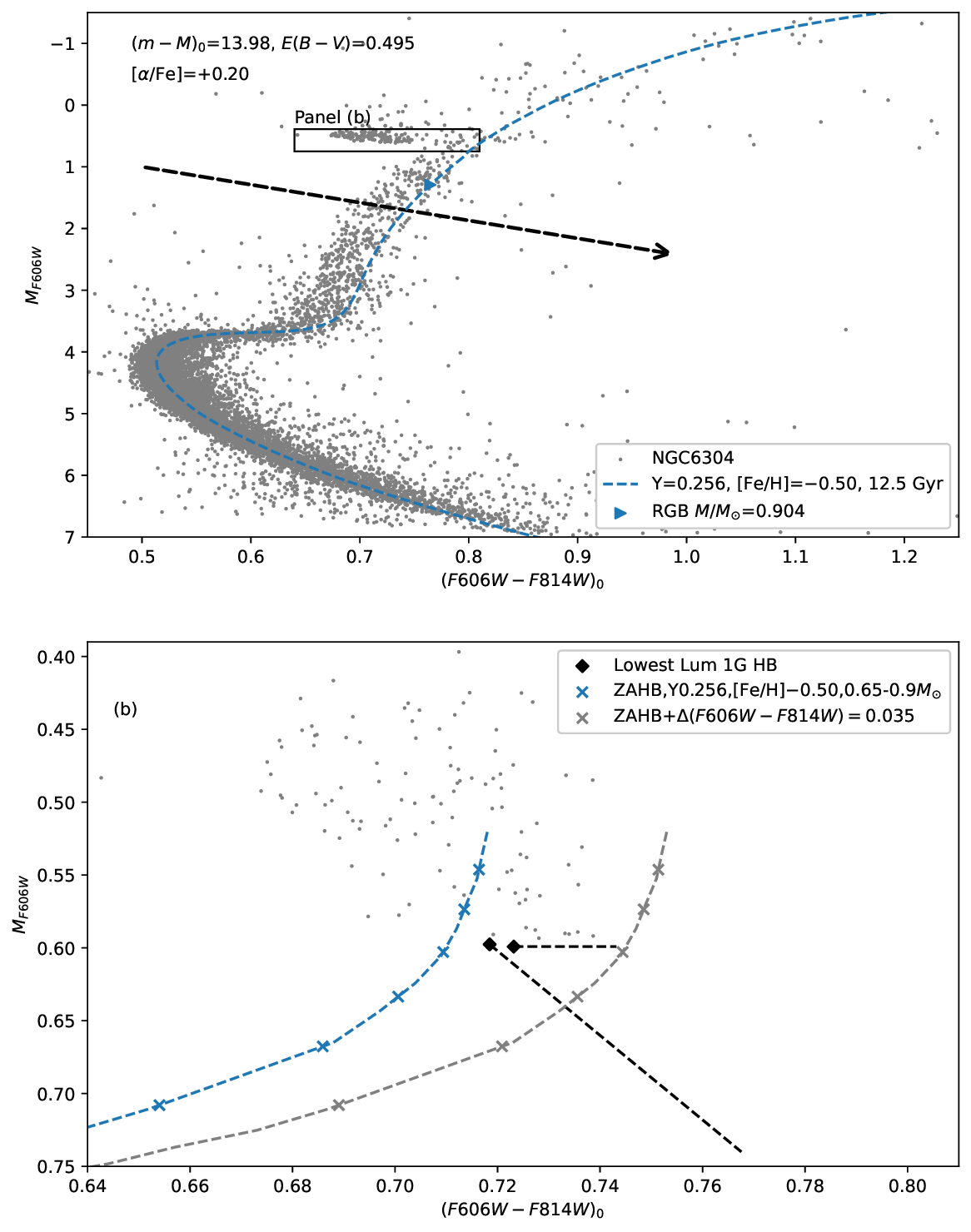}
    \caption{Colour-magnitude diagrams of NGC\,6304 over-plotted with Victoria isochrones and ZAHBs.
    Similar to Fig.~\ref{fig:NGC6352CMD}, but for NGC\,6304.
    \textit{Panel a:} Observed HST CMD compared to a model matched as described in the text. Model composition and RGB mass are given in the legend. Additional parameters are given in the top left corner. The reddening vector is shown as a black dashed arrow. Zoom box corresponding to panel (b) is marked. \textit{Panel b:} Zoom on the HB. HB stars are marked with grey points. A ZAHB is over-plotted in blue with parameters as given in the legend, and the same ZAHB shifted in colour is shown in grey. The lowest luminosity HB star is marked with a diamond and lines extending from it are used to estimate the HB mass. The sloped dashed line is along the reddening vector. Note that depending on whether one shifts horizontally or along the reddening line, different stars are in this case the least luminous one.
    }
         \label{fig:NGC6304CMD}
   \end{figure}

We repeat here the procedure for NGC\,6304 where we used $\rm [Fe/H]=-0.50$ and [$\alpha$/Fe]=$+$0.20, close to the values measured from APOGEE DR17 spectra by \citet{VACSchiavon2024}, see Table~\ref{tab:GCs}. Although they have a number of RHB stars in their sample, they are unfortunately not among the stars with HST photometry, so we cannot repeat the exact procedure from NGC\,6352 to obtain the reddening estimate. Still, they are stars from the middle to the reddest part of the HB in the $T_{\rm eff}$ range 4800-5000 K, which turn out to fall in the approximate photometric $T_{\rm eff}$ range with the reddening that we adopt. The matching is shown in Fig.~\ref{fig:NGC6304CMD}. As can be seen from panel (b), the least luminous HeCB star depends on whether it is evaluated horizontally or along the reddening line. For this cluster, the separation between 1G and 2G populations is not clear, so we have not done that. However, the RHB stars at $F606W-F814W$=0.7 and below are very likely 2G stars and therefore disregarded. Using the same procedure as for NGC\,6352, we obtained a HeCB mass in the range 0.74-0.81 $\mathrm{M_{\odot}}$ resulting in a mass loss on the RGB of $\Delta M=0.09-0.16 \,\mathrm{M_{\odot}}$, similar to what we obtained for NGC\,6352.     

\begin{table}[hbt!]
\caption{GC parameters}  
\label{tab:GCs}      
\centering                          
\begin{tabular}{l | ccl }
\hline\hline                 
GC & [Fe/H] & [$\alpha$/Fe] & Reference\\
\hline                                   
NGC\,6352 & $-0.55$ & $+0.20$ & \citet{Feltzing2009} \\
NGC\,6304 & $-0.48$ & $+0.25$ & \citet{VACSchiavon2024} \\
\hline                                   
\end{tabular}
\end{table}

If the smallest HeCB star mass was instead deduced from colour while ignoring the absolute magnitude, it could be much different. However, that would then depend strongly on the exact reddening, differential reddening, assumptions on multiple populations and their helium contents, colour-temperature relations and the temperature scale of the models. Because of the unsolved model issues that can cause significant differences in model temperatures, for example the mixing length and surface boundary conditions, it is safer to rely on absolute magnitude instead of colour (or luminosity instead of effective temperature) to estimate the minimum HeCB mass from the CMD of metal-rich globular clusters, where HeCB stars of different masses are located much closer together in colour, and the mass-discrimination therefore becomes poorer.

\FloatBarrier

\end{appendix}

\end{document}